\documentclass[fleqn,usenatbib]{mnras}

\usepackage{newtxtext,newtxmath}

\usepackage[T1]{fontenc}

\DeclareRobustCommand{\VAN}[3]{#2}
\let\VANthebibliography\thebibliography
\def\thebibliography{\DeclareRobustCommand{\VAN}[3]{##3}\VANthebibliography}

\usepackage{graphicx}	
\usepackage{amsmath}	
\usepackage{tablefootnote}
\usepackage{threeparttable}
\usepackage{float}
\usepackage[colorinlistoftodos]{todonotes}
\usepackage{subfigure}

\usepackage{url}

\title[{\it AstroSat} observations of IGR J17591--2342]{\textit{AstroSat} timing and spectral analysis of the accretion-powered millisecond X-ray pulsar IGR J17591--2342}
\author[A. Singh et al.]{Akshay Singh$^{1,2}$\thanks{E-mail: akshay.singh@biu.ac.il},
Andrea Sanna$^{3\thanks{andrea.sanna@dsf.unica.it}},$
Sudip Bhattacharyya$^{4\thanks{sudip@tifr.res.in}},$
Sudip Chakraborty$^{5},$
Sarita Jangle$^{2},$
\newauthor
Tilak Katoch$^{4}$,
H. M. Antia$^{6}$  and Nitinkumar Bijewar$^{2}$
\\
$^{1}$Department of Physics, Bar-Ilan University, Ramat-Gan, 5290002, Israel \\
$^{2}$University Department of Physics, University of Mumbai, Mumbai 400098, India \\
$^{3}$Universit\'a degli Studi di Cagliari, Dipartimento di Fisica, SP Monserrato-Sestu, KM 0.7, 09042 Monserrato, Italy\\
$^{4}$Department of Astronomy and Astrophysics, Tata Institute of Fundamental Research, Mumbai 400005, India\\
$^{5}$ Universit\'e Paris Saclay, Universit\'e Paris Cit\'e, CEA, CNRS, AIM, F-91191 Gif-sur-Yvette, France\\
$^{6}$ UM-DAE Centre of Excellence for Basic Sciences, University of Mumbai, Kalina, Mumbai 400098, India\\
}

\date{Accepted 2024 November 15. Received 2024 November 15; in original form 2023 September 28}

\pubyear{2015}

\begin{document}
\label{firstpage}
\pagerange{\pageref{firstpage}--\pageref{lastpage}}
\maketitle

\begin{abstract}
 IGR J17591--2342, a transient accretion-powered millisecond X-ray pulsar, was discovered during its 2018 outburst. Here, we present a timing and spectral analysis of the source using {\it AstroSat} data of the same outburst. From the timing analysis, we obtain updated values of binary orbital parameters,  which reveal an average pulsar spin frequency of 527.4256984(8) Hz. The pulse profiles can be fit well with four harmonically related sinusoidal components with fractional amplitudes of fundamental and second, third, and fourth harmonics as $\sim13$\%, $\sim$6\%, $\sim$0.9\%, $\sim$0.2\%, respectively. The energy-dependent study of pulse profiles in the range of $3-20$ keV shows that the fractional amplitude of both the fundamental and first overtone is consistent with being constant across the considered energy band. Besides, a decaying trend has been observed for both the fundamental and first overtone in the phase-delay versus energy relation resulting in soft X-ray (2.8-3.3 keV) phase lags of $\sim$0.05 and $\sim$0.13 with respect to $\leq 15$ keV photons, for the fundamental and first overtone, respectively.
The combined spectra from the Large Area X-ray Proportional Counters and the Soft X-ray Telescope aboard {\it AstroSat} in the $1-18$ keV range can be fit well with an absorbed model consisting of a Comptonization, a blackbody and a Gaussian emission line component yielding as best-fit parameters a blackbody seed photon temperature $kT_{\rm bb}$ $\sim 0.95 \pm 0.03$  keV, and an electron temperature $kT_{\rm e}$ $\sim 1.54 \pm0.03$ keV. The spectral aspects suggest the scattering of photons from the accretion disc or the neutron star's surface.
\end{abstract}



\section{Introduction}
Accretion-powered millisecond X-ray pulsars (AMXPs) are a subclass of low-mass X-ray binaries (LMXBs) that exhibit X-ray pulsations during their outburst phases at frequencies higher than 30 Hz \citep{wijnands1998millisecond,patruno2012accreting,di2020accretion}. AMXPs typically have a magnetic field of $\sim 10^7-10^9$ G \citep{mukherjee2015magnetic}.
In LMXBs, the accretion of matter from the companion star towards the neutron star (NS) takes place through Roche lobe overflow, and depending on the accretion rate ($\dot M$), the accreting matter could fall onto the magnetic polar caps, which results in the formation of hot spots on NS surface. 
These fast-spinning NS systems are typically found in binary systems with a low-mass ($\leq$ 1 $M_{\odot}$) companion star. AMXPs often exhibit intermittent outbursts with X-ray luminosity ranging from $\sim 10^{36} - 10^{38}$ ergs s$^{-1}$.

A large number of these X-ray binary systems have been discovered in Globular Clusters, primarily in old star clusters. It is also well known that these NS belong to old systems in order for their magnetic fields to decay from their initial high strength in young NS, which is usually above $10^{12}$ G.
Later, it was proposed that these old NS are spun up to millisecond periods by matter accretion and angular momentum exchange during a LMXB stage of their evolution; this is known as the recycling scenario of pulsars \citep{bhattacharya1991formation}.   
The rotation-power mechanism is activated once these accreting systems reach a high spin frequency, even if their magnetic field, which plays an important role in their evolution, has decayed significantly. As a result, such a NS system should be observable as a rotation-powered millisecond pulsar \citep{di2020accretion}. AMXPs provide a pivotal role as the link between the accreting low mass X-ray binary systems and the rotation-powered millisecond pulsars. As long as accretion is taking place, the system remains an AMXP. However, once accretion ceases, the rotation-powered pulsar mechanism is expected to quickly reactivate. These sources are referred to as transitional millisecond pulsars (MSPs).

IGR J17591--2342 is an AMXP discovered by the International Gamma-Ray Astrophysics Laboratory (INTEGRAL) on August 10, 2018, during a Galactic Center scan \citep{ducci2018integral}. The source, however, was already active since July 22, 2018, according to an archival search in the Neil Gehrels Swift Observatory (Swift) Burst Telescope data \citep{krimm2018swift}.
The presence of a hard spectral state was revealed in the X-ray spectrum observed during its outburst in 2018 \citep{Ray+18, Gusinskia+20}. IGR J17591--2342 also exhibited a type-I thermonuclear
X-ray burst during its 2018 outburst \citep{sanna2018nustar}, which is an intriguing feature of NS X-ray binary systems. 

The spectrum of the source has been described using the emission from the accretion disc (bbodyrad in Xspec), a thermal Comptonization of seed photons in the corona (nthcomp in Xspec), and a Gaussian component to model the iron emission line at 6.4 keV generated via disk reflection \citep{sanna2018nustar,kuiper2020high}.  The work by \citet{kuiper2020high} reported a value of the hydrogen column density as (2.25$\pm$0.05) $\times $ 10$^{22}$ cm$^{-2}$ while studying the high-energy characteristics of the source. 
Recently, \citet{Manca+22} did an extensive spectral analysis of the source using data from several X-ray missions. They observed that besides the multi-colour disc blackbody and the Comptonization components, a broad Fe K$\alpha$ and other spectral line features, such as O VIII and an absorption edge at 0.87~keV, are present in a number of observations. Additionally, it was discovered that spectral evolution occurred during the outburst of the source, with blackbody emission occurring close to the blackbody radius of (3.3 $\pm$1.5)~km, while the power law index of the Comptonization model varied between 1.7 and 2.1.

From the analysis of the type-I X-ray burst, \citet{kuiper2020high} also estimated the distance of the source to be (d = 7.6 $\pm$ 0.7) kpc. From the timing of the coherent signal at $\sim$527 Hz, it has been possible to determine the orbital parameters of the binary system. According to previous studies on timing analysis of the source using epoch folding techniques by \citet{sanna2018nustar, sanna2020timing,kuiper2020high} of the {\it NICER}/{\it NuSTAR} data of the source, the pulse profile of the source can be well described by four sinusoidal harmonically related components that correspond to the fundamental, second and third harmonics and have fractional amplitudes of $\sim$ 11\%, $\sim$ 4.8\%, $\sim$ 0.7\%,  $\sim$ 0.16\%, respectively. 
The amplitude and phase lag analysis of the source showed that the pulse fraction for the fundamental component shows an increase in the lower energy level up to $\sim$ 5~keV and remains constant after that. However, the second harmonic shows a very little increase in amplitude and basically remains constant.
From 1.5 to 10 keV, the energy-dependent phase lag is roughly 200 $\mu$s \citep[corresponding to $\sim 0.1$ pulse cycles;][]{sanna2020timing, kuiper2020high}.

We present the findings from {\it AstroSat} observations of IGR J17591--2342 taken during its outburst in August 2018 to undertake spectral and temporal analyses. In section 2, we describe the procedure for the {\it AstroSat}/LAXPC  and {\it AstroSat}/SXT data analysis, and in section 3, we will discuss the results of the timing and spectral analyses and finally discuss our findings in section 4.

\begin{table*}
    \centering
    \caption{Log of X-ray observations of the source IGR J17591-2342 from {\it AstroSat} payloads.The second row shows orbits during ObsID 9000002320 when the source shows a high count rate. }
    \begin{tabular}{c|c|c|c|c||c|c|}\hline\hline
         Instrument & OBS ID & Start Time & Stop Time & Mode & Exposure & Effective Exposure \\
          & & (yy-mm-dd hh:mm:ss) & (yy-mm-dd hh:mm:ss) & & (ks) & (ks) \\\hline
         LAXPC & 9000002320 & 2018-08-23 01:03:51 & 2018-08-24 00:35:00 &  Event & 34.7 & $\sim$6.7 \vspace{0.2cm} \\
         LAXPC (Flaring) & 9000002320 & 2018-08-23   02:05:10 & 2018-08-23  22:05:51  &  Event & 28 & $\sim$28 \vspace{0.2cm} \\
LAXPC & 9000002332 & 2018-08-27 00:00:02 & 2018-08-28 02:24:40 & Event & 37.7 & 37.7 \vspace{0.2cm} \\
SXT   & 9000002320 & 2018-08-23 01:03:59  & 2018-08-24 00:07:58 & PC & 20.7 & 20.7\vspace{0.2cm} \\
SXT (Flaring)   & 9000002320 & 2018-08-23 01:03:59  & 2018-08-24 00:07:58 & PC & 20.7 & 0.0 \vspace{0.2cm} \\
SXT   & 9000002332 & 2018-08-27 00:07:01 & 2018-08-28 02:24:39& PC & 18.7 & 18.7
\vspace{0.2cm} \\
\hline\hline
\end{tabular}
    \label{tab:Obs-log}
\end{table*}

\section{OBSERVATIONS AND DATA ANALYSIS}

{\it AstroSat} is India’s first dedicated astronomy space mission \citep{agrawal2006broad,singh2014astrosat}. It was launched on September 28, 2015, and it is currently still operational. It has contributed significantly to the simultaneous investigation of celestial sources in the X-ray, optical, and UV spectral bands. Its payloads are limited to the optical and X-ray regimes (0.3 keV to 100 keV) and cover the ultraviolet (near and far). {\it AstroSat} is equipped with five payloads, including the Soft X-ray Telescope (SXT) (0.3-8 keV), three co-aligned Large Area X-ray Proportional Counters (LAXPCs) (3-80 keV), the Cadmium-Zinc-Telluride Imager (CZTI) (10-100 keV), the Ultra-Violet Imaging Telescope (UVIT), and the Scanning Sky Monitor (SSM) (2.5-10 keV), which make it a well-suited X-ray (imaging) observatory.
In this work, we have used only data from the LAXPCs and SXT because the sensitivity of the CZTI instrument was too low to yield useful scientific results.

\subsection{{\it AstroSat}/LAXPC}

The Large Area X-ray Proportional Counters

(LAXPC) are one of the major payloads on {\it AstroSat}. The LAXPC10, LAXPC20, and LAXPC30 are three co-aligned identical proportional counters that operate in the energy range of 3--80 keV with a large area of around 6000 cm$^2$ and a field of view of approximately $0.9^{\circ}$ $\times$ $0.9^{\circ}$ for all LAXPCs. Each LAXPC detector has five layers, each with 12 detector cells  \citep{antia2017calibration,yadav2016large}. It is an ideal instrument for timing analysis since the detected photons' arrival times can be recorded with a temporal resolution of 10 $\mu$s. LAXPC observed the source twice, first from 01:03 UT on August 23, 2018, to 00:35 UT on August 24, 2018, for a total exposure time of $\sim$ 34.7 ks and then from 00:02 UT on August 27, 2018, to 02:24 UT on August 28, 2018, for a total exposure time of $\sim$ 37.7 ks. The log of X-ray observations of the source IGR J17591-2342 is given in Table~\ref{tab:Obs-log}. LAXPC data were collected in Event Mode (EA), which contains information about the time, channel number, and anode ID for each detected event. We did not use the data from LAXPC10 and LAXPC30 due to leak and gain stability issues. We process level1 data with the LAXPC software version 3.4.3
, which extracts the event files, light curves, spectra, and the background for good time intervals (GTI). The background of LAXPC is evaluated from the blank sky observations  \citep{antia2017calibration}. We use the response matrix file lx20cshm01v1.0.rmf for channel-to-energy conversion to obtain the spectra and light curves in different energy ranges for performing timing and spectral analysis of the data. 

The LAXPC photon arrival times were corrected to solar system barycentric times using the AS1BARY tool provided by the {\it AstroSat} support centre adopting the JPL DE405 solar system ephemeris and source position RA (J2000) $=17^h59^m02.86^s$ and Dec $=-23^{\circ}43'08.30"$ reported in \citet{Russell+18}.

\subsection{{\it AstroSat}/SXT}

The Soft X-ray Telescope
(SXT) onboard {\it AstroSat} is one of the primary instruments sensitive to soft X-rays, which has a CCD camera and studies cosmic sources in 0.3--8 keV. For both observations, SXT was working in the Photon Counting (PC) mode with a time resolution of $\approx 2.4$ s. In PC mode, the entire $40'\times40'$ CCD detector is used for the observation of the source \citep{singh2014astrosat,singh2016orbit,singh2017soft}.

SXT observed IGR J17591--2342 on 23 August 2018 for a total exposure time of $\sim$ 20.7 ks (ObsID 9000002320), and then on 27 August 2018 for a total exposure time of $\sim$ 18.7 ks (ObsID 9000002332)  (See Table \ref{tab:Obs-log}). The level 1 data were processed by AS1SXTLevel2-1.4b pipeline to produce level 2 cleaned event files, and these files were further merged with the SXTMerger.jl 

a tool written in the language Julia. Since AstroSat data are downloaded for every orbit, individual datasets may have overlapping event lists that must be identified, rejected, and combined. The SXTMerger utility reads the level2 event lists, bad pixel lists, and Good Time Intervals (GTIs) from separate orbit event files, then checks for overlapping event data, stores only unique events, and combines the event lists. The tools generate a merged event file compatible with ftools XSELECT. The merged event files were used to extract light curves and spectrum using XSELECT 2.4, which is provided as part of the software HEASOFT 6.24. Since about 95\% of the photons are detected in an area of around $15'$, a circular region of $15'$ around the source is selected.  A pile-up correction is not required because the source count rate was low ($\approx 4.22$ cts/s). For spectral analysis, we use the background spectrum Sky-Bkg\_comb\_ELCLRd16p0\_v01.pha and response matrix file sxt\_pc\_mat\_g0to12.rmf and ancillary response file sxt\_pc\_excl00\_v04\_20190608.arf. All of these files and software are provided by the SXT POC team.

\section{Results}

\begin{figure}
	\includegraphics[width=\columnwidth]{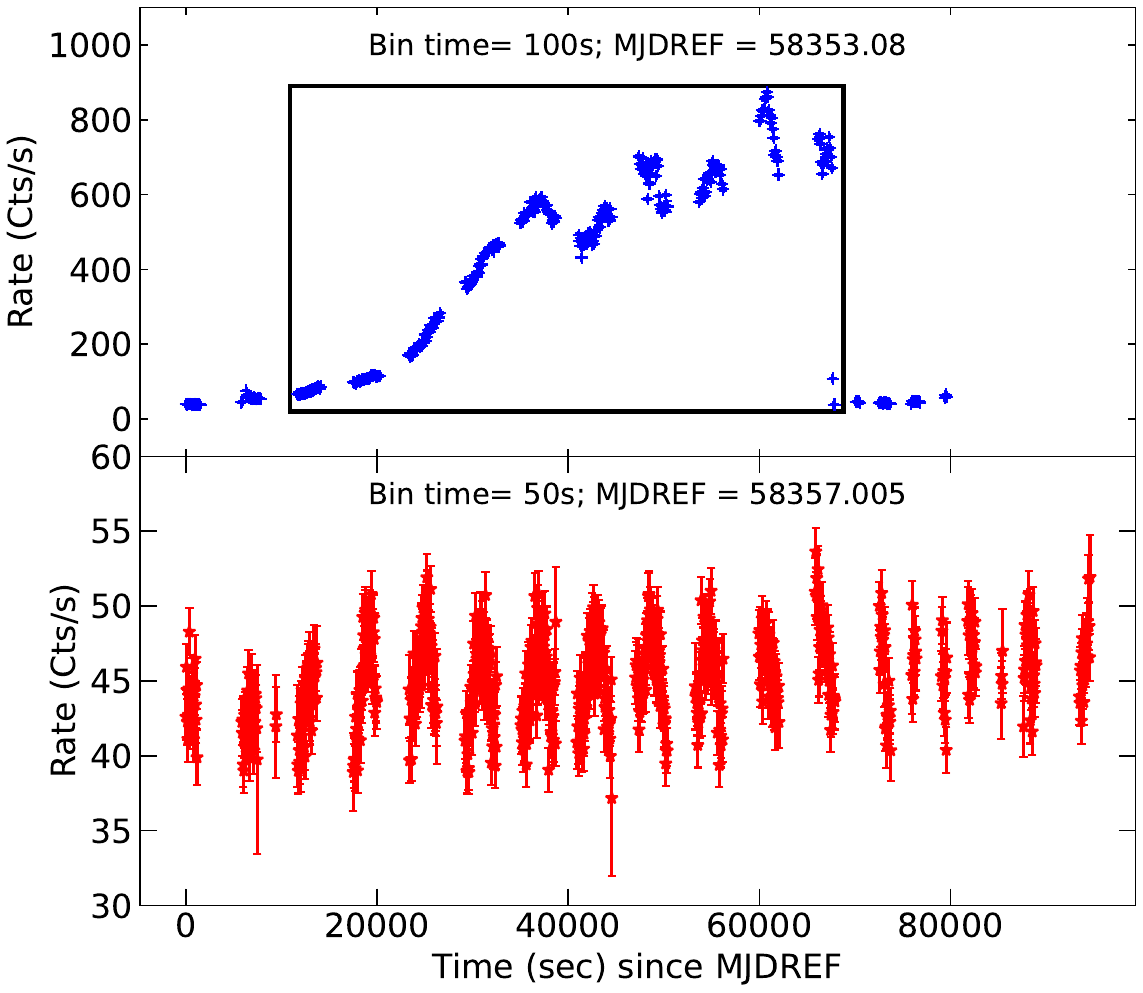}
	\caption{\textbf{Upper Panel}: The light curve of IGR J17591--2342 in the energy range 3-80 keV} as observed by {\it AstroSat}/LAXPC during its first observation on August 23, 2018. The black rectangular box here shows the time period during which there was a constant increase in the count rate of the source, which we decided to discard from further analysis. See section \ref{Light curve} for more details. \textbf{Lower Panel:} The light curve of the source during the second observation on August 28, 2018. 
    \label{fig:LightCurve}
\end{figure}

\begin{figure}
	\includegraphics[width=\columnwidth]{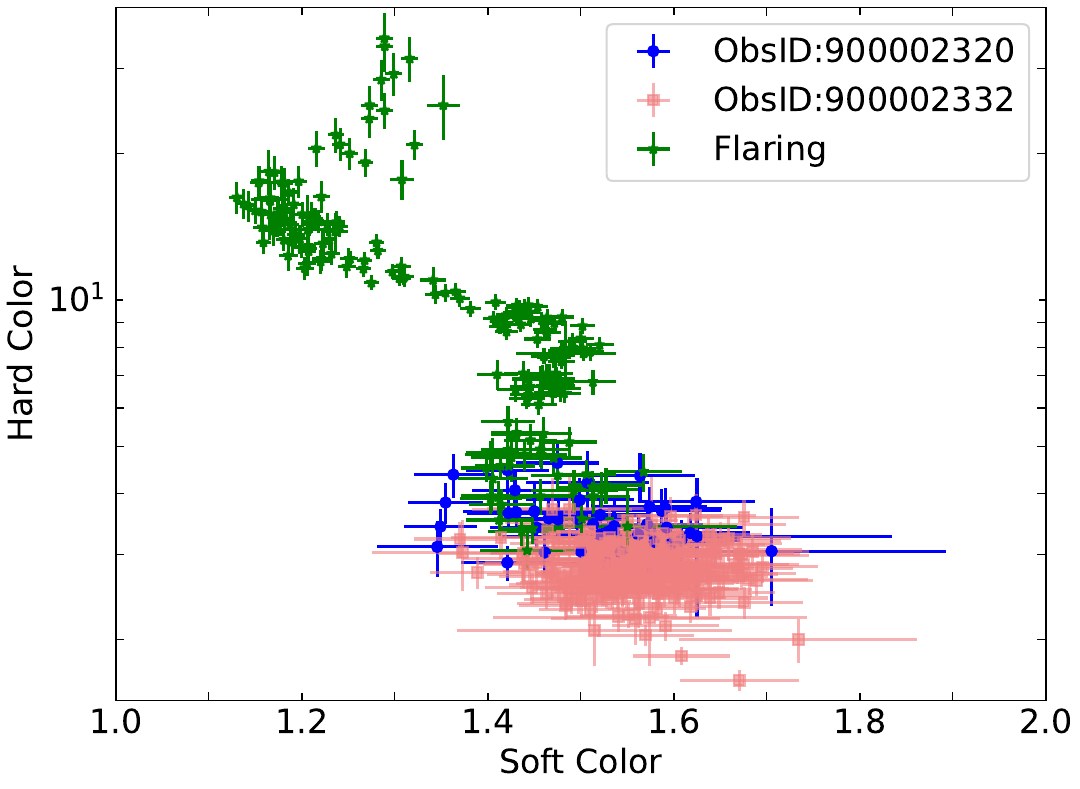}
    \caption{The colour-colour diagram from IGR J17591-2342 count rates in 4 different independent energy bands (see section \ref{colour-colour diagram} for definition) using data from {\it AstroSat}/LAXPC observations ObsID9000002320 (blue coloured), including data points from the flaring orbits (green coloured) and ObsID900000232 (light coral coloured). }
    \label{fig:Colour}
\end{figure}

\subsection{Light curve}  \label{Light curve}
The background-corrected light curve of IGR J17591--2342 observed with LAXPC onboard {\it AstroSat} is shown in Figure~\ref{fig:LightCurve}. 
A type-I X-ray burst was detected during the first observation.
This burst is not analyzed in the current paper.
Since the {\it AstroSat} is a low orbit satellite, the LAXPC light curve shows the persistent emission separated by data gaps due to Earth occultation and South Atlantic Anomaly (SAA) passages, which were then filtered out from the good time interval to obtain the persistent light curve of the source. The black rectangle in Figure~\ref{fig:LightCurve} highlights the
contaminated light curve due to the presence of the source GX 5-1 in the field of view of IGR J17591-2342 caused by a pointing issue during the observation. 

During the second LAXPC observation (ObsID 9000002332), which was performed after 4 days, the count rate suggests no further contamination from X-ray sources within the field of view. No significant changes were observed in the hardness ratio obtained by taking the ratio of count rates in the two energy bands 3–10 keV and 10–30 keV during the observations. For the second observation, we used the whole dataset for further analysis. Fig. \ref{fig:LightCurve} also shows in the bottom panel the light curve of the source during the second observation. 
The ratio of hard/soft count rates showed that the source was in the hard spectral state, and no state transition was observed from the first observation to the end of the second.

\subsection{Colour-Colour Diagram} \label{colour-colour diagram}

 A recent study by \citet{Manca+22} found that the LAXPC 20 spectrum extracted from ObsID 9000002320 has a significant mismatch with the spectrum of the source obtained from various nearby observations collected with different instruments, which they attributed to poor instrument calibration. Therefore, we opted to reconsider this observation and looked at it more rigorously. 
 Figure \ref{fig:LightCurve} shows that a few thousand seconds after the start of the observation, the source count started increasing almost to the very end of the observation. Therefore we extracted the data of high count-rate orbits (hereafter flaring orbits) and decided to plot the colour-colour diagram of the source using the data of both observations as well as the flaring orbits.  The colour–colour diagram for the source was obtained by plotting  hard colour (in logscale for visual clarity) with respect to soft colour, and it is shown in Fig. ~\ref{fig:Colour}. The soft colour is defined as
the ratio of 5–8~keV to 3–5~keV count rates, whereas hard colour is the ratio of 8–15~keV to 15–25~keV count rates. Since 25 keV is one of the highest energy for which we can extract the events with very high count statistics, we decided to generate a colour profile only up to 25~keV. The values of the soft colour range from $\sim$(1.1–1.8), whereas the hard
colour ranges from $\sim$(1.65–34.60); however, most of the soft colour lies in
the range $\sim$ (1.4–1.75), and the corresponding hard colour lies in the range $\sim$ (2.0–4.0).

The source reached its maximum luminosity within a few hours from the beginning of the observation, and then the luminosity started decreasing; however, the hard and soft colours remained the same. The soft colour increases sharply during the observation of the source and then falls back to the lower values. The colour profile of ObsID 9000002320 (removing flaring orbits) and ObsID9000002332 shows clear overlap with each other, whereas the colour plot of the flaring orbits, as shown by the green dots in figure \ref{fig:Colour}, shows a completely different behaviour.

It can be clearly observed that the colour profile changed drastically during the flaring orbits, with emission dominated in high-energy ranges. To understand the cause behind it, we tried to verify the source pointing and found that the {\it AstroSat} pointing had drifted by about 1 degree. This caused the inclusion in the field of view of a bright source, GX 5-1, which is 1.5 degrees from the source, likely explaining the sudden increase in the count rate. 
Unfortunately, the {\it AstroSat} orbit files do not give the correct information about the pointing direction. Therefore, it may be that the pointing drifts for some time until the star sensor becomes active and detects the drift. After that, it initiates corrective action, and the pointing is restored. Thus, if we look at the offset in the \textit{mkf} file, it shows correct pointing almost all the time, except for a sudden glitch in between. Normally, these glitches are about 0.1-0.2 deg, but in this case, it is about 1 degree. Thus, it appears that the pointing had been drifting for quite some time, several hours, which is why the drift appears extended. Therefore, we conclude that the satellite pointing drifted to GX 5-1, and we decided to remove the data of flaring orbits from all of the further analysis. 
After removing the flaring orbits data, the colour plot of ObsID 9000002320 indicates that the source shows very little variability, which suggests that it remained in the same spectral state throughout the observation.

\subsection{Timing analysis}
\begin{table*}
\centering
\caption{Orbital parameters and spin frequency of IGR J17591--2342 with uncertainties on the last digits reported at $1\sigma$ confidence level. {\it NICER} parameters are taken from (\citet{sanna2020timing}). The Solar System ephemeris used in the barycentering process is JPL DE405. The {\it AstroSat}/LAXPC parameters are reported for the timing analysis performed on the combined data of ObsID 9000002332 and non-flaring orbits of the ObsID 9000002320.} 

\begin{tabular}{|l|l|l|}
\hline \hline
Parameters &  {\it NICER} & {\it AstroSat}/LAXPC\\
\hline
R.A. (J2000)                                 & \multicolumn{2}{c}{$17^h\,59^m\,02^s.86$ $\pm 0.04^s$}    \\
Decl. (J2000)                                & \multicolumn{2}{c}{$-23^{\circ}43'08.30"$ $\pm$ 0.1"}    \\ \hline
Orbital period P$_{\text{orb}}$(s)                      & 31684.7503(5)        & 31684.73(1)               \\ \hline
Projected semi-major axis asin(i/c) (lt-s) & 1.227714(4)          & 1.22772(1)                \\ \hline
Ascending node passage T$_{\text{NOD}}$ (MJD)            & 58345.1719781(9)     & 58345.171980(4)           \\ \hline
Eccentricity (e)                             & $ <5\times10^{-5}$     & $ <7\times10^{-5}$   \\ \hline
$T_0$ (MJD)                                     & 58344.0               & 58344.0                   \\ 
&(58344.182-58345.532)               &  58344.0 (58344.058-58345.12) \\ \hline
&   Fundamental frequency and 1st derivative  & \\ 
$\nu_0$ (Hz)                                          & 527.425700578(9)  & 527.4256984(8)  \\ 
$\dot{\nu}$ (Hz $s^{-1}$)                                     & $-7.4(4)\times10^{-14}$  & (1.9 $\pm$ 0.8) $\times10^{-12}$         \\ \hline

$\chi^2$/d.o.f.                                    & 876.4/355            & 112.5/61              \\ \hline \hline
\end{tabular}
\label{tab:timing}
\end{table*}


\begin{figure}
	\includegraphics[width=\columnwidth]{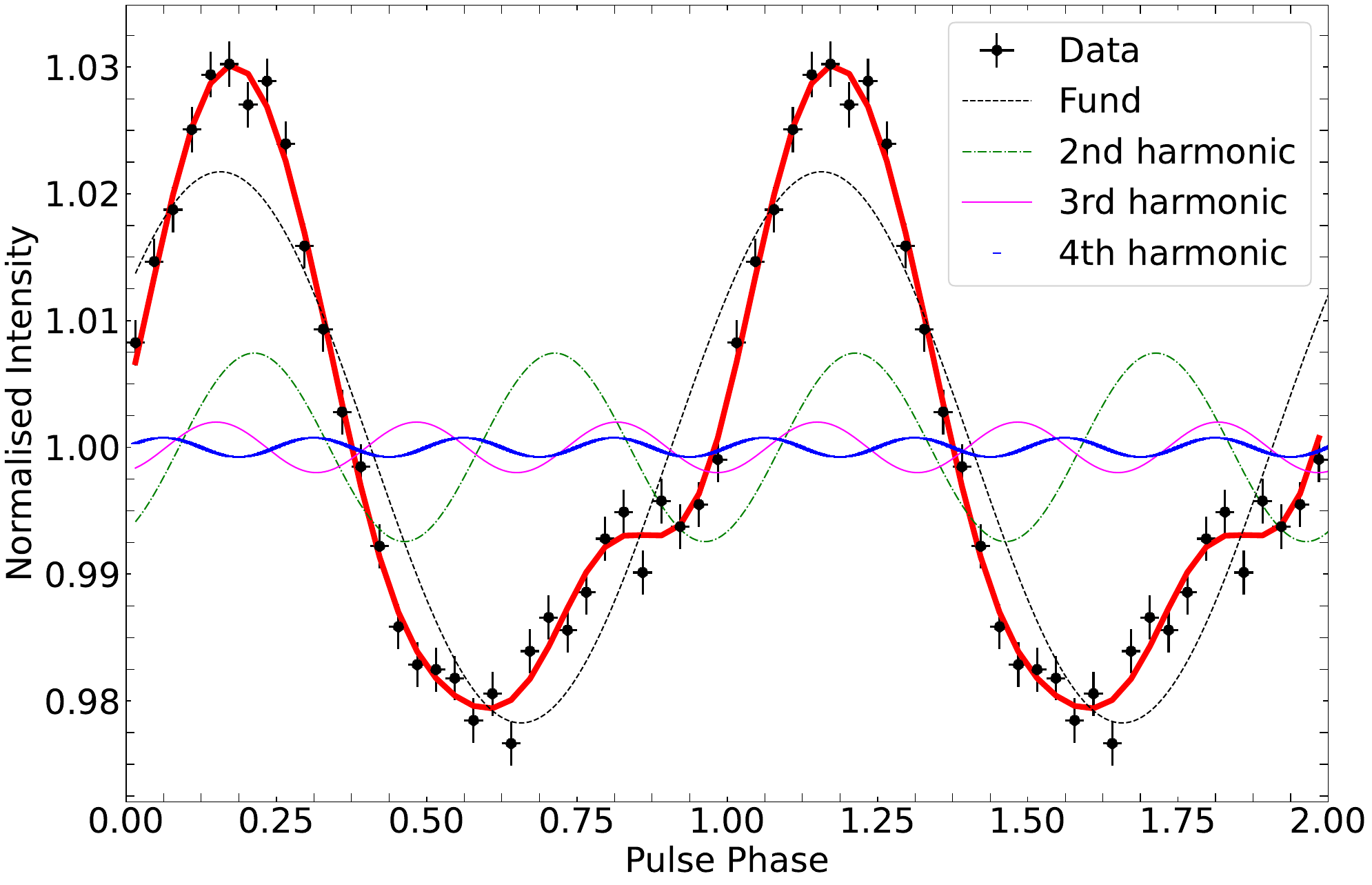}
    \caption{IGR J17591--2342 pulse profile using 32 bins (black points) in the energy range of 3-20 keV from epoch-folding the {\it AstroSat} observations after correcting for the orbital parameters reported in Table \ref{tab:timing}. The best-fitting model (red line) is the superposition of four sinusoidal functions with harmonically related periods. Here, we show two cycles of the pulse profile for better clarity.}
    \label{fig:PSD}
\end{figure}

Considering the timing solution listed in \citet{sanna2020timing} as initial parameters, we investigate possible updates on the orbital parameters by using the {\it AstroSat} datasets.
We started by correcting the photon time of arrivals for the binary orbital motion using the expression 
\begin{equation}
    \dfrac{z(t)}{c} = \dfrac{a \sin{(i)}}{c} \left[\sin\left(\dfrac{2 \pi}{P_\text{{orb}}}(t - T_{\text{NOD}})\right)\right],
\end{equation}
where t is given by a recursive relation 
\begin{equation}
    t_{em} \simeq t_{arr} -  \dfrac{z(t_{arr})}{c},
\end{equation}
where $t_\text{{em}}$ is photon emission time, $t_\text{{arr}}$ is the photon arrival time at the Solar system barycentre, $z(t_{em})$ represents the projection, along the line of sight, of the separation between the NS and the barycentre of the binary system, $c$ is the speed of light, $a \sin{(i)}/c$ is the projected semi-major axis of the NS orbit in light seconds, $P_\text{{orb}}$ is the orbital period and $T_{\text{NOD}}$ is the time of passing the ascending node \citep[for more details on the method, see][]{sanna2016timing}. We performed the barycentric correction of LAXPC photon arrival times with the BARYCORR tool using the JPL DE-405 Solar system ephemeris. For the timing analysis, we used the combined data of LAXPC ObsID 9000002332 plus the non-flaring part of ObsID 9000002320. We conducted an epoch-folding analysis of data to investigate the presence of pulsations. This analysis involved the utilisation of 16 phase bins, and an initial spin frequency denoted as $\nu_0$ obtained from \citep{sanna2020timing}, which had a value of 527.4257000578 Hz. This corresponds to the spin frequency that was originally measured when the source was first identified during its outburst in 2018. To systematically explore a range of possible spin frequencies in the vicinity of $\nu_0$, we performed a search that was carried out by incrementing the frequency in steps of $10^{-8}$ Hz, with a total number of 10001 steps. This approach allowed us to investigate a range of nearby frequencies to assess whether the observed data exhibited pulsations at different frequencies beyond the initial value of $\nu_0$. The corresponding average pulse profile in the energy range of 3-20 keV is well-fitted using four sinusoidal functions harmonically related, with the fundamental, first, second and third overtones having a fractional amplitude of $\sim$ 13 $\pm$ 3\%, $\sim$ 6 $\pm$ 1.9\%, $\sim $0.9 $\pm$ 0.28\%, $\sim $0.2 $\pm$ 0.06\% respectively as seen in Fig. \ref{fig:PSD}. In order to perform phase-connected timing analysis to investigate the NS ephemeris, we split the data into $\sim$ 1000 s intervals and epoch-folded into 8 phase bins, finding the mean spin frequency $\bar{\nu}$= 527.4256984 Hz with respect to the epoch $T_0$ = 58344.0 MJD.

We modelled the temporal evolution of the pulse phase delays
with the relation:
\begin{equation}
 \Delta \phi(t) = \phi_0+ \Delta \nu_0 (t - T_0 ) + \dfrac{1}{2}\dot{\nu}(t -T_0)^2 + R_\text{{orb}(t)},
\end{equation}

where $T_{0}$ stands for the reference epoch for the timing solution, $\dot{\nu}$ is the spin frequency derivative,  $\Delta\nu_0 = (\nu_0 - \bar{\nu})$ is the difference between the frequency at the reference epoch and the spin frequency employed for epoch-folding the data, and $R_\text{{orb}(t)}$ represents the phase residual, which arises due to disparities between the accurate set of orbital parameters and those utilized to adjust the arrival times of photons \citep{deeter1981pulse}. Through precise timing analysis of {\it AstroSat} observations, a spin-up frequency derivative of roughly $(1.9 \pm 0.8) \times 10^{-12}$ Hz s$^{-1}$ was obtained.
We find that the timing analysis performed using the {\it NICER} dataset by \citet{sanna2020timing} and \citet{kuiper2020high} and our derived parameters are consistent within uncertainties. Table~\ref{tab:timing} displays the best-fitting parameters that were obtained from our analysis as well as the from \citet{sanna2020timing}.

\subsubsection{Energy dependence of pulse profile} \label{energy-dependence}
\begin{figure*}
  \centering
  \subfigure{\includegraphics[scale=0.23]{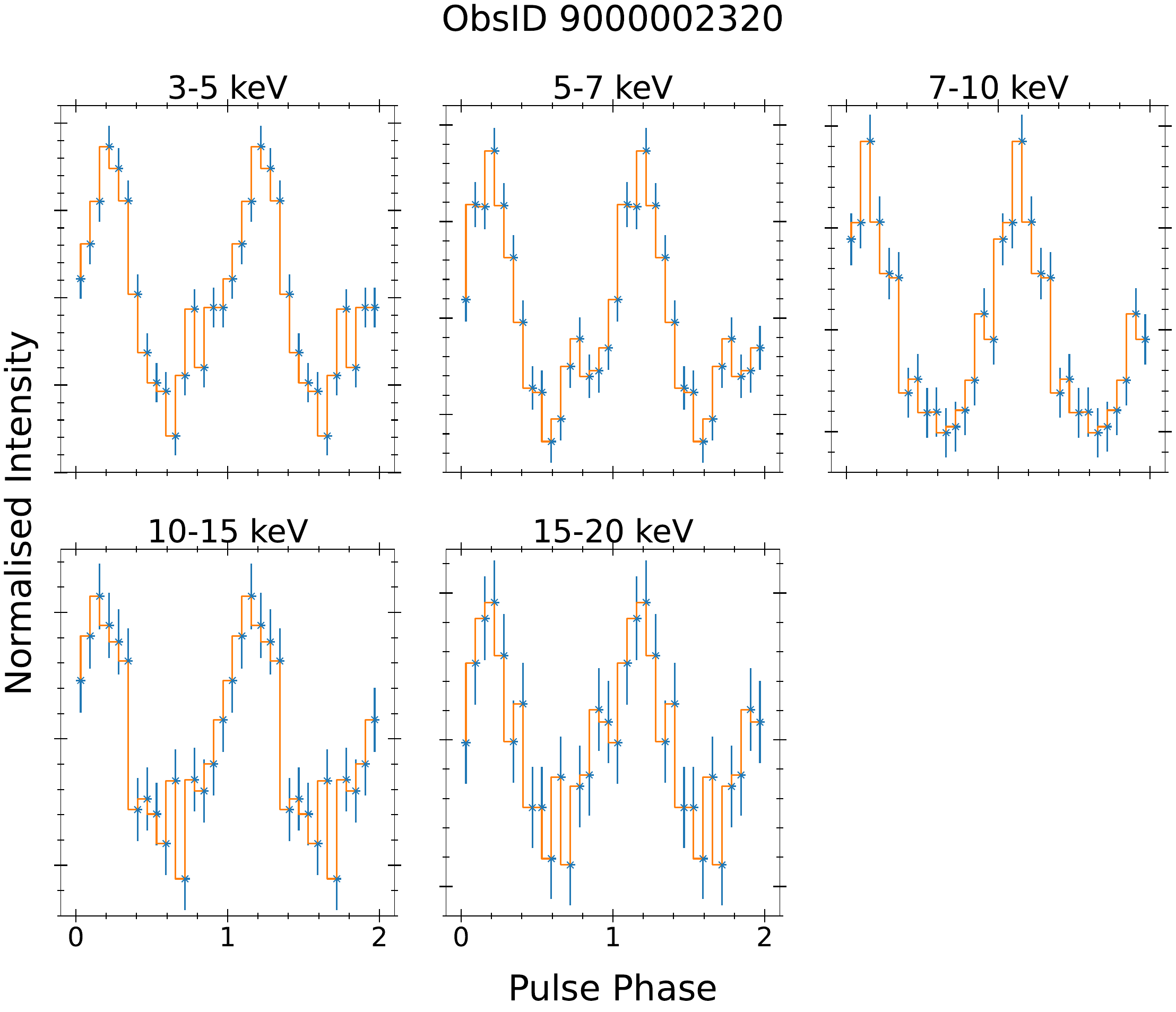}}\quad
  \subfigure{\includegraphics[scale=0.23]{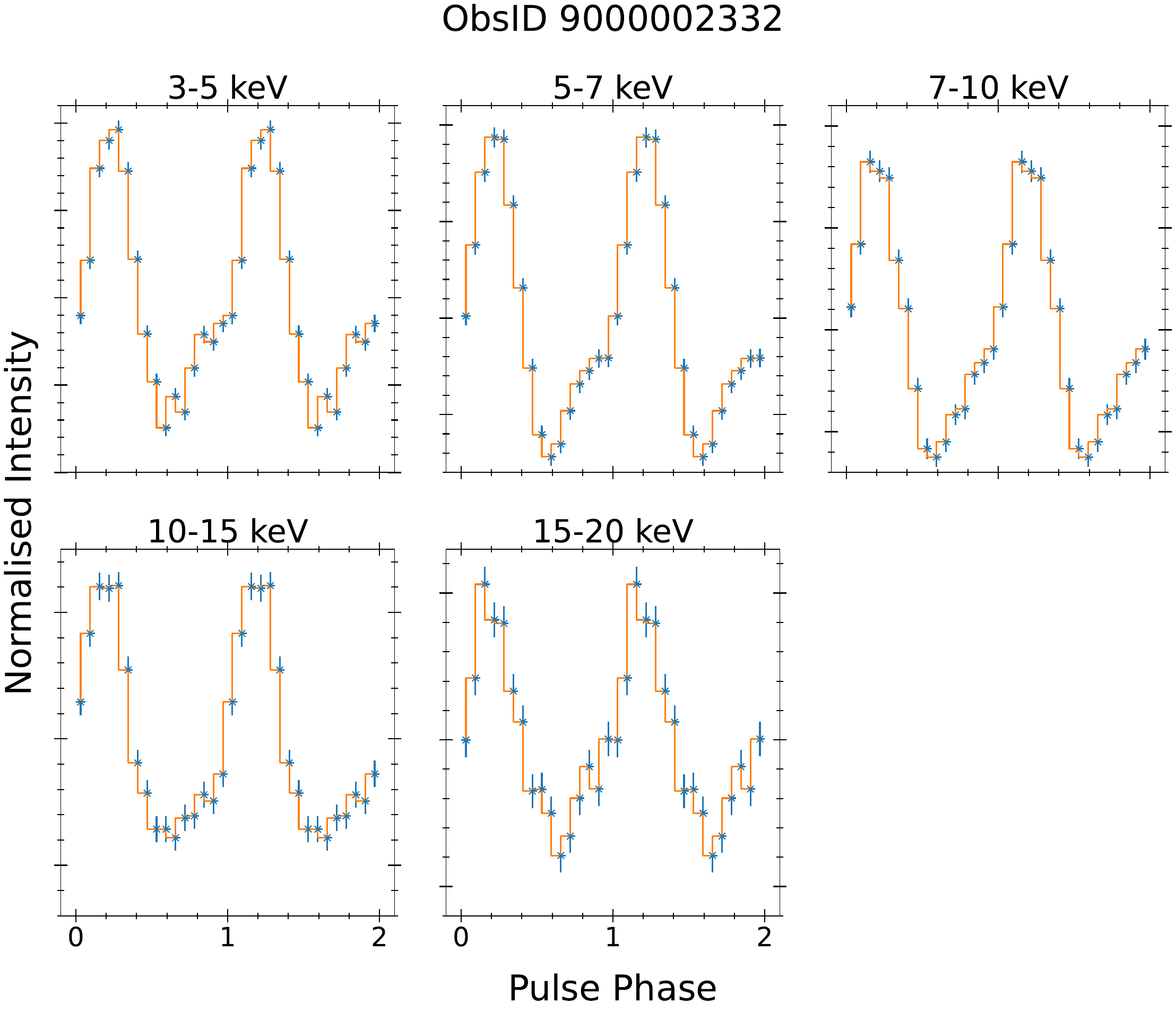}}
  \caption{Pulse profiles (16 bins) as a function of energy for IGR J17591-2342 using data from {\it AstroSat}/LAXPC ObsID9000002320 (left panel) and ObsID9000002332 (right panel) across the 3-20 keV energy range. Data points are aqua-coloured, while fits are in orange. The intensity axis is rescaled for each energy bin, but for comparison purposes, it remains the same for a given energy band for both observations. The pulse profile shapes are consistent with those shown in \citet{kuiper2020high}}
\label{fig:Energy_profile}
\end{figure*}

We investigated the energy dependency of the average pulse profile from {\it AstroSat} event data by dividing the energy range from 3~keV to 20~keV into 19 smaller bins of varying size to detect the pulsation in particular energy bands and comparing the pulse morphology as a function of energy.
Figure~\ref{fig:Energy_profile} shows the energy-dependent pulse profile for five large energy bands from 3-20 keV for both of the LAXPC observations for visual representation. It was observed that all of the profiles peak at around phase 0.25. The amplitude of the pulse decreases with the energy range. It is noted that the timing analysis was limited only up to 20~keV because, in the higher energy ranges, the pulse profiles become less than 3 sigma significant due to the decrease in the count rate at higher energies. Due to the reduced exposure time because of pointing drifts in ObsId 9000002320, the error bars on the amplitudes are higher compared to the other observation. The intensity axis is rescaled independently for each energy bin. However, for making meaningful comparisons, the intensity axis retains a consistent scaling within a specific energy bin for both observations. We found that the amplitude and the shape of pulse profiles are consistent with the pulse profiles reported in
\citet{kuiper2020high}.

\begin{figure}
	\includegraphics[width=\columnwidth]{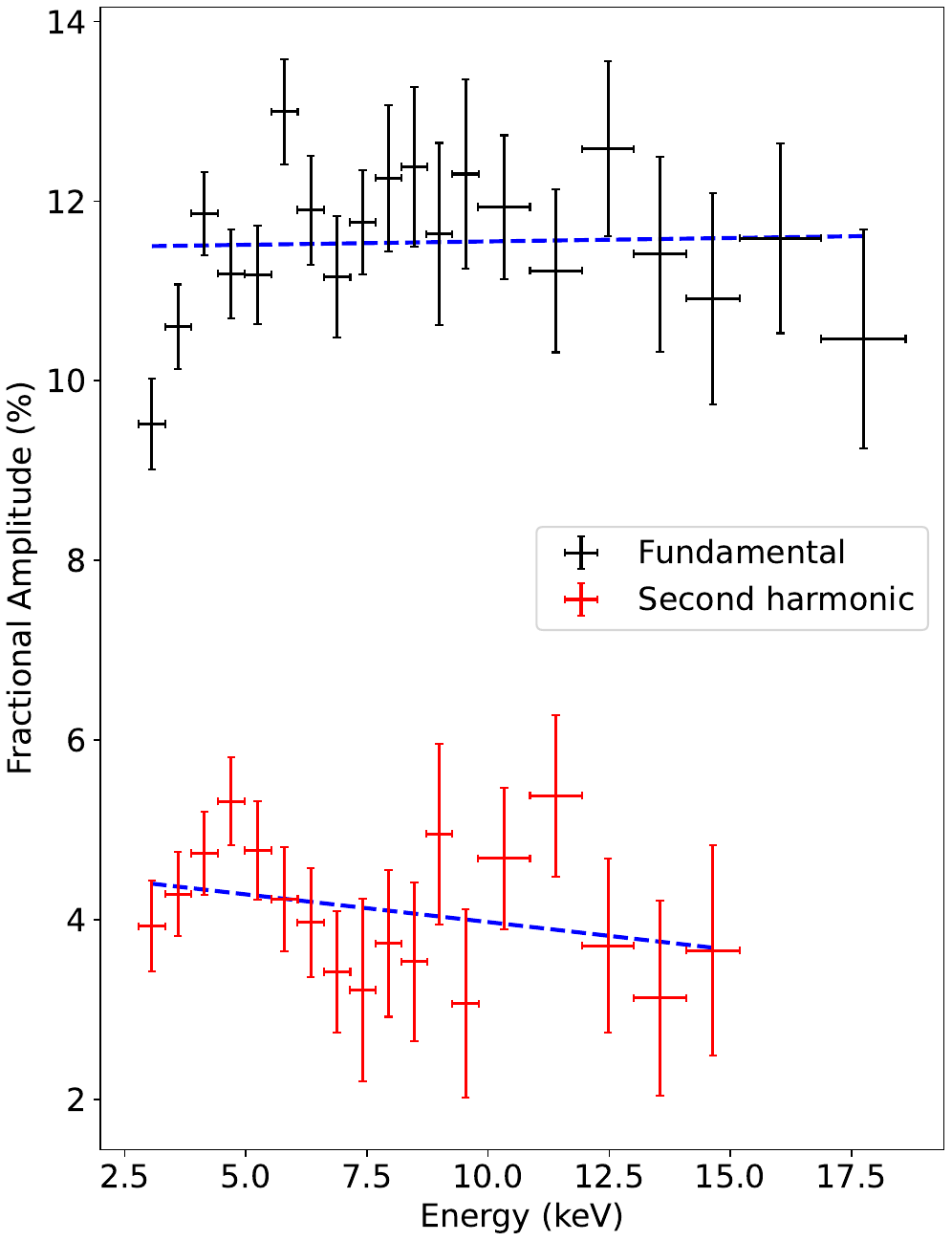}
    \caption{The pulse fractional amplitude of the fundamental component and its first overtone plotted as a function of energy using background corrected data from ObsID 9000002332. The blue dotted lines represent linear model fits to} the fundamental and first overtone data,
respectively.
    \label{fig:Fractional-amplitude}
\end{figure}

For the epoch-folding analysis, we used the data only from ObsdID 9000002332 since the data from ObsID 9000002320 gives poor statistics due to the reduced exposure time. Each epoch-folded pulse profile was modelled with two harmonically related sinusoidal functions of the unitary period to determine the corresponding amplitude and fractional part of the phase residual.
For the analysis, only folded profiles for which the ratio between the amplitude and its 1 $\sigma$ uncertainty is equal to or larger than three were considered. 
The result of this analysis is shown in Figure ~\ref{fig:Fractional-amplitude}. From the figure, we found no particular trend in the data. We fitted the data with a linear model but F-test revealed that the overall trend in the fractional amplitude remains constant with energy.

\begin{figure}
	\includegraphics[width=\columnwidth]{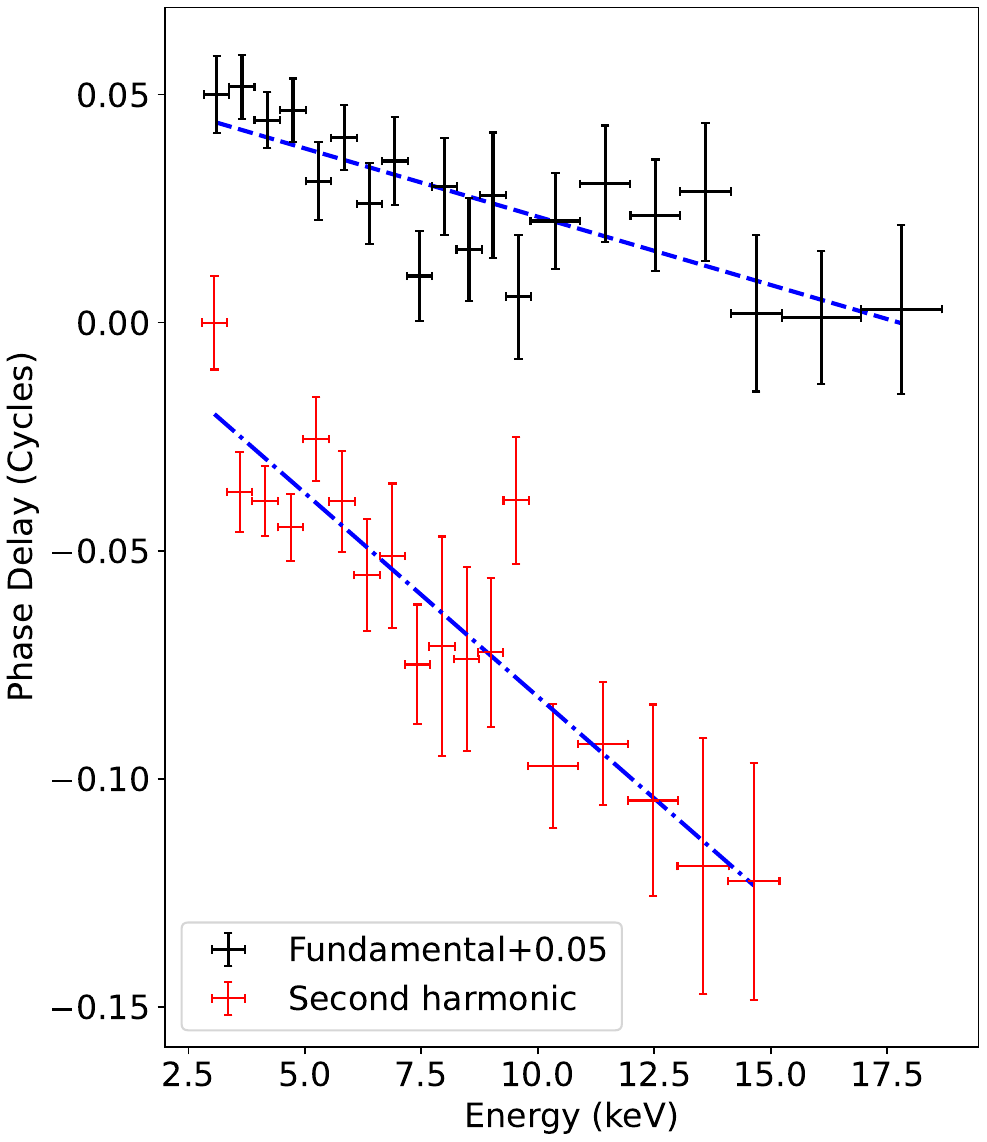}
        \caption{Phase lags plotted as a function of energy, obtained from the fundamental component (black points) and the first harmonic (red points) with respect to the first energy band (2.8-3.3 keV) for LAXPC ObsID9000002332. A constant shift of 0.05 is applied to the fundamental component to visualize data points better. The blue dotted lines represent linear model fits to} the fundamental and first overtone data, respectively.
    \label{fig:Phase lag}
\end{figure}

Figure ~\ref{fig:Phase lag} shows the energy-dependent pulse arrival phase delay relative to band 1. The fundamental component is shifted by adding a constant value of 0.05 for a better representation of the result.
The soft lag observed up to 5~keV is of the order of 0.03 phase cycles, and the pulsation in the highest energy band (17.5 keV) leads the band 1 (2.8-3.3~keV) by 0.13 cycles. To understand the trend in the phase delay, here also we fit a simple linear model to the fundamental component, which shows a linear increase in the phase delay with the energy. A similar behaviour was also observed for the second harmonic, which shows that the phase delay increases by $\sim$0.13 cycles up to 15 keV. 

The second harmonic for the fractional amplitude as well as the phase lag, has been plotted only up to 15 keV because they become non-statistically significant (less than 3$\sigma$ significance) at the high-energy ranges. However, the overall phase decay remains consistent with the results of \citet{kuiper2020high,sanna2020timing,Kaho+20}.

\subsection{Spectral Analysis}
We performed the spectral analysis of the source for the two observations 
from the LAXPC and SXT instruments onboard {\it AstroSat}. For the spectral fitting, we only used the LAXPC20 detector from the LAXPC instrument; the data from LAXPC10 and LAXPC30 detectors were excluded due to instrument calibration issues at higher energies. The spectra for the LAXPC were extracted only for the top layer due to the source being relatively faint. LAXPC20  spectrum for ObsID 9000002320 was extracted by removing the thermonuclear burst and the flaring part. We used the response file lx20cshm01v1.0.rmf and added 2\% systematics \citep{Mishra+17} for the spectral fitting.

For extra assurance that the pointing drift did not change the spectrum, we extracted the spectra of ObsID 9000002320 by removing the flaring part, fitted it with the same model as ObsID 9000002332, and found that the spectral parameters agreed quite well. As a result, we conclude that there was no calibration issue and that the suspected increase in intensity was caused by the satellite drifting towards a nearby bright source, GX 5-1. 
The SXT image of the source taken from ObsID 9000002320 revealed a
brightening on one edge of the image, which can also be attributed to the satellite pointing drift. Therefore, we tried to extract the spectrum by filtering GTI to remove the brightened part or to take a small area to avoid the brightening; the source being very faint, we did not get enough count statistics to generate the spectrum for this observation; therefore, we decided to discard this observation for spectral analysis. For ObsID 9000002332, we extracted the photon-counting mode Level 2 data for SXT and reduced the data using SXT software. We extracted the image of the source using the orbit-merged event file and applied a circular region of $15'$ radius centred on the source for further extraction of the source spectrum using HEASoft package XSELECT. We used the files sxt\_pc\_mat\_g0to12.rmf and SkyBkg\_comb\_EL3p5\_Cl\_Rd16p0\_v01.pha, and SXT software as provided by SXT POC, as the energy redistribution and background spectral files, respectively.  Finally, we used sxt\_pc\_excl00\_v04\_20190608.arf, corrected for the vignetting effect using the SXTARFModule tool, as our effective area file.

\begin{table*}
	\centering
	\caption{The obtained best fit spectral parameters for IGR J17591--2342 for {\it AstroSat} LAXPC/SXT with uncertainties at 1$\sigma$ confidence level.}
	\label{tab:example_table}
	\begin{tabular}{lcccr} 
		\hline \hline
		Model & Parameter & LAXPC & LAXPC+SXT & Combined LAXPC/SXT \\
             &  &   ( ObsID 2320) & (ObsID 2332) & (ObsID 2320 + ObsID 2332) \\
		\hline
		TBabs & $N_{H} (10^{22} \text{cm}^{-2})$ & 2.09 (fixed) & 2.09 (fixed) & 2.09 (fixed)\\ \hline
		Bbodyrad & $kT_{\rm bb}$ (keV) & 0.66$\pm$ 0.042 & 0.93$\pm$ 0.02 & 0.95$\pm$ 0.03\\ 
		&   Norm \text($ km^2/D^2_{10} \text{kpc}^2$)  &  65 $\pm$ 31 &  12.06 $\pm$ 1.30 &  12.01 $\pm$ 1.02\\ \hline
		Nthcomp & $\Gamma$ & 1.68$\pm$ 0.05  & 1.53 $\pm$ 0.04 & 1.54 $\pm$ 0.03\\ 
		&   $kT_{\rm seed}$ (keV) &0.66$\pm$ 0.042 & 0.93$\pm$ 0.02 & 0.95$\pm$ 0.03\\ 
		&   $kT_{\rm e}$ (keV)    & 40 (fixed)   &   40 (fixed)  &   40 (fixed) \\ 
		&   Norm (\text{Photons} $\text{cm}^{-2}\text{s}^{-1} \text{keV}^{-1}$) & <1.02 $\times 10^{-3}$ & < 4.35 $\times 10^{-3}$ & < 4.28 $\times 10^{-3}$   \\ \hline
		Gauss   &   LineE (keV) & 6.41 $\pm$ 0.15 & 6.49 $\pm$ 0.35  & 6.43 $\pm$ 0.05\\
		&   Sigma (keV) &   1.10 $\pm$ 0.14&   1.24 $\pm$ 0.17 &   1.27 $\pm$ 0.11 \\ 
        &   Norm (\text{Photons} $\text{cm}^{-2}\text{s}^{-1}$) &   < 4.356 $\times 10^{-3}$ &   < 1.762 $\times 10^{-3}$ &   < 2.017 $\times 10^{-3}$ \\ \hline
		Cross-calibration constant    & -  & 1 (fixed) &  1.07$\pm$0.02  & - \\ 
        (SXT/LAXPC20) &  & &  \\ \hline
		Unabs. Flux  & $F_{0.1-100 \hspace{0.08cm} {\rm keV}}$ &  4.22$\times  10^{-10}$ & 4.62 $\times  10^{-10}$ \\ 
        & (erg  cm$^{-2}$ s$^{-1})$ & & \\ \hline
	         $\chi^2$/dof &  & 22.37/16 & 211.17/188 & 280.48/209 \\ \hline
		\hline
	\end{tabular}
	\label{tab:spectrum}
\end{table*}

We used data from persistent emission obtained with LAXPC (5-18 keV) and SXT (1--7~keV) to perform spectroscopy. Considering the intrinsic resolution of the LAXPC instrument, the LAXPC20 spectra were rebinned to have at least 25 counts per bin using the tool grppha. The energy range was limited to only 18 keV because, at higher energy levels (18-30 keV), the background intensity becomes more prominent and overwhelms the count rates from the source. 

Similarly, the SXT spectrum was binned at a minimum of 25 counts per energy channel.  First, we tried to fit the spectrum using the compPS model, which was observed to give a better fit for the persistent spectrum of the source \citep{kuiper2020high}, but it did not fit well for our data; therefore, we used the same models as the latest work by \citet{Manca+22}. Therefore we fitted the continuum X-ray spectrum of IGR J17591--2342 for both of the observations within the spectral fitting package XSPEC 12.12 \citep{1996ASPC..101...17A} using the model \textsc{Tbabs*(nthcomp+bbodyrad+gauss)}, where \textsc{Tbabs} is the interstellar absorption model for the neutral hydrogen density, which we kept constant at $2.09 \times 10^{22}$ $\rm{cm^{-2}}$, adapted from the previous analyses \citet{kuiper2020high,Manca+22} with the abundances set to Wilms \citep{2000ApJ...542..914W} and the cross-section set to Vern \citep{1996ApJ...465..487V}.  The \textsc{bbodyrad} component is used to model the blackbody-like component of the spectrum. The thermally Comptonized emission was described using the \textsc{nthcomp} model, which takes into account the emission caused by the thermal distribution of electrons that Compton up-scatter the soft X-ray seed photons. The parameter {\it int-type} and {\it redshift} were fixed to 0. The parameters {\it int-type} equal to 0 means that the seed photons are coming from the {\it bbody}-like component. We tied the blackbody temperature $kT$ with the seed photon temperature $kT_{bb}$ of the nthcomp model and let the other four parameters free to vary. A constant component was also added for LAXPC20/SXT cross-calibration uncertainties while fitting both spectra simultaneously.

The continuum spectral fit showed a clear bump in the residuals around $\sim$ 5-7 keV, likely representing  Fe K$\alpha$ emission, which we modelled using a Gaussian component. We observed that the Gaussian profile has a width larger than 1 keV; therefore, we restrict the value of the centroid energy of the Gaussian component to be between 6-7 keV. However, the error on the width of the Gaussian profile was about 1.1~keV, giving a broad Gaussian line profile.  The model for the combined fit of all the datasets gives the best spectral fits with the reduced $\chi^2$ value 1.34 (209 d.o.f.).
We determined the radius of NS ($R_{bb}$) from bbodyrad normalization ($k_{bb}$) using  $R_{bb} = \sqrt{k_{bb}}D_{10}$, where $D_{10}$ indicates the distance to the source in the units of 10 kpc under the assumption that the bbodyrad component describes only photons coming from the NS surface. However, we emphasize that this may not always be true. Assuming the distance to the source at $d = (7.6 \pm 0.7)$~kpc \citep{kuiper2020high}, the blackbody radius we calculated was $ 12.10\pm 1.12$~km, compatible with the whole NS surface emitting the soft radiation.
Our spectral study revealed that the overall photon index of the spectrum was $1.54\pm0.03$, and the blackbody temperature was ($0.95 \pm 0.03$)~keV.

However, the other parameters were in good agreement with the literature \citep{Manca+22, sanna2018nustar, kuiper2020high}. We report the average unabsorbed flux of $4.82\times 10^{-10}$~erg~cm$^{-2}$~s$^{-1}$ in 0.1–100 keV energy range of {\it AstroSat} SXT/LAXPC band.
Table~\ref{tab:spectrum} shows the best fit spectral parameters obtained at a 1$\sigma$ confidence level for both observations.

\begin{figure}
	\includegraphics[width=\columnwidth]{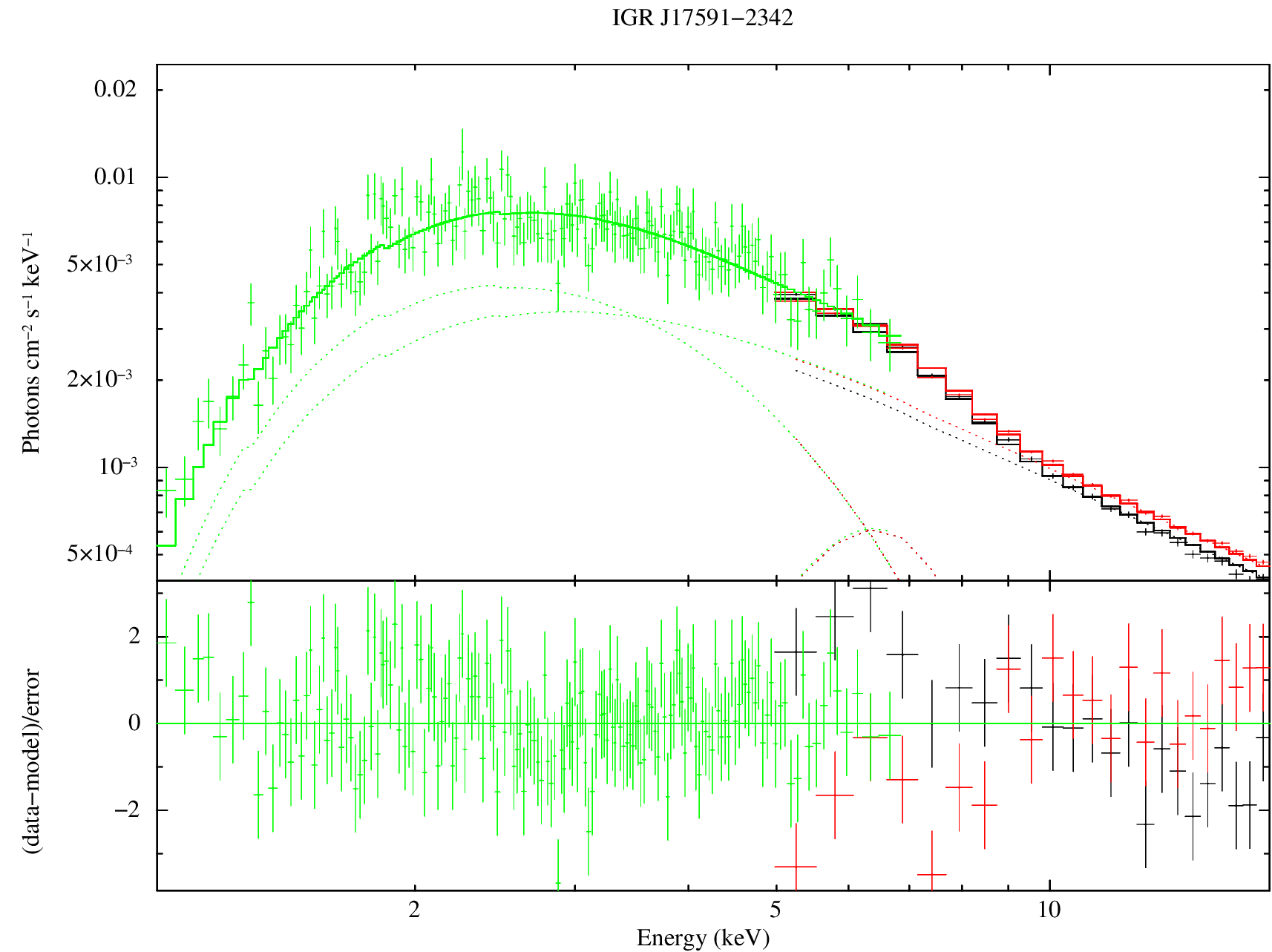}
    \caption{\textbf{Top Panel} - Best fitted time-averaged spectra of LAXPC and SXT modelled with \textsc{Tbabs*(constant*nthcomp+bbodyrad+gauss)}. The LAXPC/9000002320 spectrum is shown in black colour, and LAXPC/9000002332 and SXT/9000002332 are shown in red and green colours, respectively. The nthcomp, gaussian component and bbody model components are shown as dashed lines. \textbf{Lower panel} - Residuals from the best-fit model in units of standard deviations.}
    \label{fig:Spectrum}
\end{figure}

\section{Discussion and Conclusions}
IGR J17591--2342 is an accreting millisecond X-ray pulsar discovered in outburst by {\it INTEGRAL} in 2018. We have analysed the {\it AstroSat}/LAXPC and SXT data to investigate its spectral and temporal properties during its 2018 outburst. {\it AstroSat} observed the source on two separate days, for a total of $\sim$ 73 ks. Our analysis suggests no change in the spectral state of the source observed during the observations.
Also, we observed that in the dataset, there was an increase in intensity likely associated with the contamination of a bright X-ray source in the field of view of the instrument due to large satellite pointing drifts, which significantly altered the spectra of the source. Therefore, we decided to remove the contaminated time interval to obtain a more realistic spectrum of the source.

Epoch folding the data using the X-ray ephemeris reported in the literature confirmed the presence of coherent X-ray pulsations at $\sim527$~Hz.
Phase-coherent timing analysis of the collected events from {\it AstroSat} enabled us to 
estimate an orbital solution of the source, which gave compatible results within errors with the solution reported for the 2018 outburst of IGR J17591--234 \citep{sanna2020timing,kuiper2020high}.
The pulse profile obtained in the energy range below 15 keV shows that the fundamental pulse fractional amplitude
remains nearly constant. 
An increasing behaviour was also observed in the \citet{sanna2020timing} using {\it NICER} data and \citet{kuiper2020high} using {\it NuSTAR}, {\it NICER} and {\it XMM-Newton} data. The timing analysis of the data revealed no significant spin frequency derivative, $(1.9 \pm 0.8) \times 10^{-12}$ Hz/s, during the time interval investigated. 
Assuming a source distance of 7.6 kpc, and a neutron star mass of $1.4~M_{\odot}$, and a radius of $\sim$12 km that we measured from blackbody normalization.
Furthermore, we considered the possibility of the accretion disk terminating at the co-rotation radius. Our calculations indicated a maximum anticipated spin-up rate, or its derivative, in the neutron star's rotation, which would be in the order of a few $10^{-13}$~Hz/s. 

AMXPs usually show energy-dependent pulse profiles and fractional amplitudes \citep{patruno2012accreting}. According to \citet{Patruno+10}, certain AMXPs exhibit an intricate fluctuation of fractional amplitude, with rises and declines in their value seen for various energies. The findings of \citet{patruno2012accreting} suggested that as the energy rises, the blackbody might increase its flux, and the pulse fraction increases accordingly.  
AMXPs tend to also exhibit a characteristic behaviour known as ``soft lags''. Soft lags manifest as a consistent temporal delay in the arrival times of pulsations originating from softer energy bands when compared to those emanating from harder energy bands. This temporal disparity in pulse arrival times provides valuable insights into the pulsar's emission processes and the interplay between different energy components in the observed X-ray emissions. The origin of the phase lags has been discussed in terms of two different models. \citet{2003MNRAS.343.1301P} presented a two-component model, a soft blackbody component and a Compotonized component from Compton up-scattering of the cooler seed photons by the hot electrons. They showed that the blackbody and the Comptonized components have different angular distributions and are affected by the Doppler effects in a different way.

The blackbody component dominates the low energy range. With the increase in the energy of soft photons, the Comptonized component becomes more dominant, and the pulsation in the high energy range leads to that of the lower energy ranges, causing the soft lag. Later, \citet{2007ApJ...661.1084F} also proposed a Comptonization model for the energy-dependent soft/hard time delays and and pulsed fractional amplitudes found for the NS pulsed emission. By hard X-ray photons' down-scattering in the comparatively cool plasma of the disc or NS surface, they could explain the soft delays. Their model assumes that the hot electrons in the accretion column upscatter some of the soft X-ray photons that are coming from the disc or NS surface. Due to the thermal Comptonization of soft photons, this effect causes hard lags in the pulse profiles. The spectral analysis of the source also suggests that in the low energy range, the blackbody component dominates over the Comptonization.  Therefore we believe that the model by \citet{2003MNRAS.343.1301P} better explains the phase lag in this source, as has already been discussed in the work by \citet{Kaho+20} for the {\it NICER} data.
 
The X-ray continuum spectrum of the IGR J17591--2342 is well described by both blackbody and Comptonized blackbody models, with a blackbody temperature of approximately 0.6 keV. Additionally, a broad K$\alpha$ line was observed in the source's spectrum, which contrasts with the very narrow lines reported by \citet{sanna2020timing} and \citet{Manca+22}. \citet{kuiper2020high} did also detect a 4.9 $\sigma$ bump near the 6.4 keV line; however, they attribute this to calibration issues or a blend of lines. However, all of the other parameters were consistent with the earlier studies of the source.

\section*{Acknowledgements}
Akshay would like to thank  Rahul Sharma for fruitful discussions regarding SXT data analysis and results and is deeply indebted to Damien B\'egu\'e for all of his assistance. SC acknowledges funding from the European Union's Horizon 2020 research and innovation program under grant agreement n$^{\circ}$101004168, the XMM2ATHENA project. This publication uses data from the {\it AstroSat} mission of ISRO, archived at the Indian Space Science Data Centre (ISSDC). We thank Ajay Ratheesh for the valuable discussion on the inclusion of CZTI data analysis. We thank the LAXPC Payload Operation Center (POC) and the SXT POC at TIFR, Mumbai, for providing the necessary software tools.


\section*{Data and Code Availability}


The data used in this paper are publicly available from the ISSDC, at \url{https://webapps.issdc.gov.in/astro_archive/archive/Home.jsp}, using the ObsIDs mentioned in Table~\ref{tab:Obs-log}. All the codes, software and instrument-related files like responses and gain used in this paper can be obtained from the following webpage \url{http://astrosat-ssc.iucaa.in/data\_and\_analysis}



\bibliographystyle{mnras}
\bibliography{example} 






\bsp	
\label{lastpage}
\end{document}